%% file: EXO-15-009_temp.tex
\pdfoutput=1

\documentclass[11pt,twoside,a4paper,cmspaper,final,collab]{cms-tdr}
\def\svnVersion{416685}\def\cmsCernNoTag{CERN-EP/2017-047}\def\cmsCernDate{\today}\def\cmsMessage{Published in the Journal of High Energy Physics as \href{http://dx.doi.org/10.1007/JHEP07(2017)013}{\doi{10.1007/JHEP07(2017)013.}}}

\begin{document}\cmsNoteHeader{EXO-15-009}

\hyphenation{had-ron-i-za-tion}
\hyphenation{cal-or-i-me-ter}
\hyphenation{de-vices}
\RCS$Revision: 406801 $
\RCS$HeadURL: svn+ssh://svn.cern.ch/reps/tdr2/papers/EXO-15-009/trunk/EXO-15-009.tex $
\RCS$Id: EXO-15-009.tex 406801 2017-05-27 12:30:20Z alverson $
\newlength\cmsFigWidth
\ifthenelse{\boolean{cms@external}}{\setlength\cmsFigWidth{0.85\columnwidth}}{\setlength\cmsFigWidth{0.4\textwidth}}
\ifthenelse{\boolean{cms@external}}{\providecommand{\cmsLeft}{top\xspace}}{\providecommand{\cmsLeft}{left\xspace}}
\ifthenelse{\boolean{cms@external}}{\providecommand{\cmsRight}{bottom\xspace}}{\providecommand{\cmsRight}{right\xspace}}
\newcommand{\CHI}{\ensuremath{\chi_{\text{dijet}}}\xspace}
\newcommand{\intlumi} {\ensuremath{2.6\ \textrm{fb}^{-1}}\xspace}
\newcommand{\mjj} {\ensuremath{M_{\textrm{jj}}}\xspace}
\newcommand{\ptave}{\ensuremath{\langle p_\textrm{T}  \rangle}\xspace}
\newcommand{\QBHMC} {{\textsc{qbh}}\xspace}

\newcolumntype{d}[1]{D{.}{.}{#1}}
\newcolumntype{+}{D{+}{\pm}}

\cmsNoteHeader{EXO-15-009}
\title{Search for new physics with dijet angular distributions in proton-proton collisions at $\sqrt{s} = 13$\TeV}

\date{\today}

\abstract{
A search is presented for extra spatial dimensions, quantum black holes, and quark contact interactions
in measurements of dijet angular distributions
in proton-proton collisions at $\sqrt{s} = 13$\TeV. The data were collected with the CMS detector at the
LHC and correspond to an integrated luminosity of 2.6\fbinv. The distributions are found to be in
agreement with predictions from perturbative quantum chromodynamics that include electroweak corrections.
Limits for different contact interaction models are obtained. In a benchmark model, valid to next-to-leading order in QCD and in which only left-handed quarks participate, quark contact interactions are excluded up to a scale of 11.5 and 14.7\TeV
for destructive or constructive interference, respectively. The production of quantum black holes is excluded for masses
below 7.8 or 5.3\TeV, depending on the model.
The lower limits for the scales of virtual graviton exchange in the Arkani-Hamed--Dimopoulos--Dvali
model of extra spatial dimensions are in the range 7.9--11.2\TeV, and are the most stringent set of limits available.
}

\hypersetup{%
pdfauthor={CMS Collaboration},%
pdftitle={Search for new physics with dijet angular distributions in proton-proton collisions at sqrt(s) = 13 TeV},%
pdfsubject={CMS},%
pdfkeywords={CMS, physics, QCD, contact interactions, extra dimensions}}

\maketitle

\section{Introduction}

In the standard model (SM), pointlike parton-parton scattering in
high energy proton-proton collisions can give rise to dijet events,
containing at least two jets with large transverse momenta (\PT).
Such events may be used to test the perturbative predictions of quantum chromodynamics (QCD) and to search for signatures of
new physics (NP), such as quark substructure or compositeness~\cite{Terazawa:1980,CI-1,CI-2}, as well as for
additional compactified large spatial dimensions~\cite{ADD1,ADD2}, and quantum black holes~\cite{QBH,QBH1,QBH2}.

The angular distribution of dijets with respect to the beam direction
is sensitive to the dynamics of the scattering process, yet is not strongly dependent on the
parton distribution functions (PDFs), since the angular distributions
of the dominant underlying processes, $\PQq\Pg\to\PQq\Pg$, $\qqbar(\PQq')\to\qqbar(\PQq')$,
and $\Pg\Pq\to\Pg\Pg$, are similar~\cite{Combridge:1983jn}.
The dijet angular distribution is typically expressed in terms of
$\CHI= \exp[\abs{(y_1 -y_2)}]$, where
$y_1$ and $y_2$ are the rapidities of the two jets with highest \pt (the leading jets).
The choice of this variable is motivated by the fact that
the \CHI distribution is uniform in Rutherford scattering, and permits
signatures from NP that have more-isotropic
scattering-angle distributions than QCD processes to be more easily
identified and examined as they could
produce an excess of events at low values of \CHI.

A common signature of quark compositeness models
is the appearance of new interactions between quark constituents
at a characteristic scale, $\Lambda$, that is much larger than the
quark masses.  At energies well below $\Lambda$, these
interactions are approximated through contact interactions (CI)
characterized by four-fermion couplings.
The most stringent limits on quark CI come from searches
studying dijet angular distributions at high dijet invariant
masses (\mjj)~\cite{ATLASDijet8TeV,dijetangularCMS8TeV,ATLAS:2015nsi},
and inclusive jet \pt distributions~\cite{jetptCMS}.
A previous search performed by the CMS Collaboration at the CERN LHC
at $\sqrt{s} = 8\TeV$
using dijet angular distributions~\cite{dijetangularCMS8TeV}
provided lower limits on $\Lambda$ ranging from 8.8 to 15.2\TeV,
for a variety of CI models.
The ATLAS Collaboration recently presented a similar analysis at $\sqrt{s} = 13\TeV$ in Ref.~\cite{Aaboud:2017yvp}, which obtained lower limits on quark CI scales in the range 13.1--29.5\TeV, depending on the model.

The Arkani-Hamed--Dimopoulos--Dvali (ADD) model~\cite{ADD1,ADD2}
of compactified large extra dimensions (ED) provides a possible
solution to the SM hierarchy problem. In proton-proton collisions
at the LHC, the ADD model predicts signatures of virtual graviton exchange
that result in a nonresonant enhancement of dijet production and an angular
distribution that differs from the QCD expectation.
Signatures from virtual graviton exchange have previously been sought at the LHC
in dilepton~\cite{Chatrchyan:2012kc,Khachatryan:2014fba,Aad:2012bsa,Aad:2014wca},
diphoton~\cite{Chatrchyan:2011jx,diphotonCMS,diphotonATLAS}, and dijet~\cite{dijetsLHC,dijetangularCMS8TeV} final states,
and the most stringent limits on the cutoff scale come from the dijet angular analysis of CMS at
$\sqrt{s} = 8$\TeV~\cite{dijetangularCMS8TeV} that range from 5.9 to 8.4\TeV, depending on the model of virtual graviton exchange.

In models with large ED, the fundamental Planck scale can
be comparable to the electroweak scale, which can make black hole production possible
at the LHC~\cite{BH1,Banks:1999gd,Emparan:2000rs,BH2,BH3}. Semiclassical black holes that have
masses much larger than the Planck scale and decay into multijets through Hawking
radiation~\cite{BH4}, have previously been sought in multijet final states~\cite{CMSBH0,CMSBH1,CMSBH2,ATLASBH1,ATLASBH2}.
Quantum black holes (QBH), produced with mass close to the reduced Planck scale, decay
predominantly into dijets that can be studied using dijet angular
distributions~\cite{QBH,QBH1,QBH2}. Recent searches for QBH in dijet final states
at the LHC are reported in Refs.~\cite{CMSBH1,CMSQBH1,CMSBH2,CMSQBH2,ATLASDijet7TeV,ATLASDijet8TeV,ATLAS:2015nsi}.
Lower bounds on QBH masses published by the CMS Collaboration at $\sqrt{s} = 8$\TeV
range from 5.0 to 6.3\TeV for different QBH models~\cite{CMSQBH2}.

Measurements of dijet angular distributions at the S{\Pp\Pap}S by the UA1
Collaboration~\cite{bib_UA1}, at the Fermilab Tevatron by the D0~\cite{bib_d0-prd-chi,bib_d0-prl-chi} and CDF
~\cite{bib_cdf-prl-chi} Collaborations, and at the LHC by the
ATLAS~\cite{bib_atlas-chi,bib_atlas-dijet2010,ATLASDijet7TeV,dijetangularATLAS8TeV,ATLAS:2015nsi,Aaboud:2017yvp}
and CMS~\cite{bib_cms-dijets,bib_cms-chi2010,bib_cms-chi2011,dijetangularCMS8TeV} Collaborations
have previously been reported. In this paper, the earlier
searches by CMS~\cite{dijetangularCMS8TeV,bib_cms-chi2011,bib_cms-chi2010}
at $\sqrt{s} = 7$ and 8\TeV are extended to higher \mjj using data that
correspond to an integrated luminosity of \intlumi at $\sqrt{s} = 13$\TeV, following the same analysis strategy reported by the previous publications.
The measurement of the dijet angular distributions, unfolded
for detector effects, is presented and is then analyzed for the presence of contact interactions,
large extra dimensions, and quantum black holes.

\section{The CMS detector and event selection}

The CMS apparatus is based on a superconducting solenoid of 6\unit{m}
internal diameter, providing an axial field of 3.8\unit{T}.  Within the
solenoid and nearest to the interaction point are the silicon pixel
and strip trackers.  Surrounding the tracker volume are the lead
tungstate crystal electromagnetic calorimeter and the
brass and scintillator hadron calorimeter.  The pixel and tracker cover a
pseudorapidity region of $\abs{\eta} < 2.5$ while the calorimeters cover $\abs{\eta} < 3.0$.
In addition, CMS has extensive forward
calorimetry, which extends the coverage to $\abs{\eta} <5.0$. Finally, muons are
measured in gas-ionization detectors embedded in the steel flux-return yoke
of the solenoid, with a coverage of $\abs{\eta} < 2.4$.
A more detailed description of the CMS detector, together with a definition
of the coordinate system used and the relevant kinematic variables, can be found in Ref.~\cite{bib_CMS}.

Events are reconstructed using the particle-flow
algorithm~\cite{pflow:2009,pflow:2010} to identify and reconstruct
individual particles from each collision by combining information from
all CMS subdetectors.  Identified particles include charged
hadrons, neutral hadrons, electrons, muons, and photons. The particles are
clustered into jets using the anti-\kt algorithm~\cite{antikt:2008}
with a jet size parameter $R =0.4$ as implemented in \FASTJET 3.0.1~\cite{Cacciari:2011ma}.
The jet energies are corrected for the
combined response function of the calorimeters using corrections
derived from data and Monte Carlo (MC) simulations~\cite{bib_jecjinst8tev}.
To compare data with next-to-leading order (NLO) and
{\PYTHIA{}~8.212}~\cite{pythia6,pythia8:2008} predictions, particle-level
jets are reconstructed by applying the same jet clustering algorithm to the four-vectors
of generated stable particles
(lifetime $c\tau > 1\cm$)
in the case of {\PYTHIA{}8}, and to the
outgoing partons in the case of NLO predictions.

A two-tiered system, consisting of the level-1 (L1) and high-level (HLT) triggers, is used
by CMS to record events of interest~\cite{Khachatryan:2016bia}.
The selection criteria for this analysis are based upon
the scalar sum of the transverse momenta of the jets reconstructed by the
L1 and HLT systems. The selection threshold was varied over the course of the
data taking and was between 100 and 175\GeV at L1 and between 650 and 800\GeV at HLT.

In the offline event selection, events with at least two reconstructed jets are selected.
Spurious jets from noise or non-colliding backgrounds are rejected by applying
loose quality criteria~\cite{bib_pfjetid} to jet properties. For each
event a reconstructed primary vertex~\cite{Chatrchyan:2014fea} is required to lie within
$\pm$24\unit{cm} of the detector center along the beam line and within
2\unit{cm} of the detector center in the plane transverse to the beam.
The primary vertex is defined as the vertex with the highest sum of squares
of all associated physics-object transverse momenta.
The physics objects are the objects returned by the anti-\kt algorithm applied to all charged tracks associated with the vertex, plus the corresponding associated missing transverse momentum.

The two leading jets are used to measure the dijet angular distributions
in several regions of \mjj which are, in units of  \TeVns,
1.9--2.4, 2.4--3.0, 3.0--3.6, 3.6--4.2, 4.2--4.8, and $>$4.8.
The highest three \mjj ranges were chosen to maximize the expected
sensitivity to the NP signals considered. The phase space for this analysis
is defined by selecting events with $1\le\chi_\text{dijet}<16$ and
$\abs{y_\text{boost}}<1.11$, where $y_\text{boost} = (1/2)(y_1 + y_2)$.
This selection restricts the rapidities $\abs{y_1}$ and $\abs{y_2}$ of
the two highest-\pt jets to be less than 2.5 and their \pt
to be larger than 200\GeV.
The trigger efficiency exceeds 99\% over the entire phase space.
The highest value of \mjj
observed in the data is 6.8\TeV.

\section{Unfolding and experimental uncertainties}

Fluctuations in jet response from the jet \pt resolution of the
detector can cause low-energy jets to be misidentified as leading jets.
Such fluctuations can produce bin-to-bin migrations in both \CHI and \mjj.
The measured distributions are corrected for these migrations and unfolded
to the particle level using the D'Agostini iteration method~\cite{D'Agostini1995487} implemented
in the \textsc{RooUnfold} package~\cite{roounfold}.  The unfolding corrections are determined
using a two-dimensional response matrix mapping the generator-level \mjj and \CHI distributions
onto the measured values. This matrix is
obtained using particle-level jets from the {\PYTHIA{}8} MC event generator that
are smeared in \pt using a double-sided Crystal-Ball parameterization~\cite{crystalball} of the response.
This parameterization takes into account the full jet energy resolution including non-Gaussian tails.
The unfolding corrections change the shape of the dijet angular distributions by less than 1\% across $\chi_\text{dijet}$ in
the lowest \mjj range, and by less than 5\% across $\chi_\text{dijet}$ in the highest \mjj
range.

The dijet angular distributions are normalized to the integrated dijet cross sections in each
\mjj range, denoted $(1/\sigma_\text{dijet})(\rd\sigma_\text{dijet}/\rd\chi_\text{dijet})$,
where $\sigma_\text{dijet}$ is the cross section in the analysis phase space considered.
The normalized angular distributions are relatively insensitive to many systematic effects.
The main systematic uncertainties come from the
jet energy scale, the jet energy resolution, and the unfolding correction. The effects of these
uncertainties on the dijet angular distributions are described below.

The maximum jet energy scale uncertainty is less than 1\%
and has a dependence on $\eta$ of less than 1\% per unit of
$\eta$~\cite{bib_jecjinst8tev,jets2015} in the phase space of the analysis. The resulting
uncertainty in the $\chi_\text{dijet}$ distributions due to the
jet energy calibration uncertainties is found to be
2.2\% in the lowest \mjj range and 3.6\% in the highest \mjj range,
over all $\chi_\text{dijet}$ bins.

The jet energy resolution uncertainty is evaluated by
changing the width of the Gaussian core of the Crystal-Ball parameterization
of the response by up to $\pm$10\%~\cite{bib_jecjinst8tev,jets2015}, depending upon
the jet $\eta$, and comparing the resultant
unfolded distributions before and after these changes. This uncertainty is found to be less than 1.1\%.
The systematic uncertainty from the modelling of the tails of the jet energy resolution
is evaluated using a Gaussian function to parameterize the response,
and assigning as an uncertainty half of the difference
between the unfolded distributions determined from this Gaussian ansatz and the nominal
correction, which covers the differences between the jet energy resolution tails in the data and simulation.
The size of this uncertainty is less than 1\%.

A source of uncertainty to the unfolding correction arises from the
use of a parameterized model to simulate the jet \pt resolution of the detector.
This uncertainty is estimated by comparing the smeared
$\chi_\text{dijet}$ distributions to the ones from a detailed
simulation of the CMS detector using \GEANTfour~\cite{Agostinelli:2002hh}, and is
found to be less than 1\% in all \mjj ranges.
An additional systematic uncertainty is evaluated to account for
mismodelling of the dijet kinematic distributions by applying
the unfolding corrections determined with {\PYTHIA{}8} to smeared
$\chi_\text{dijet}$ distributions from \MADGRAPH{}5\_a\MCATNLO 2.2.2~\cite{Alwall:2014hca},
and comparing the results with the generated $\chi_\text{dijet}$ distributions.
This uncertainty is found to be less than 1\% for all \mjj.

The effect of additional interactions in the same or adjacent
proton bunch crossings (pileup) relative to the interaction of interest
is studied by comparing \CHI distributions in simulated samples
where the distribution of pileup interactions is varied according to its uncertainty.
The effect of this variation on the \CHI distributions is observed to be less than 1\%.

A summary of the leading experimental systematic uncertainties is provided in Table~\ref{tab:sys}.
Though in the subsequent analysis
of the data the uncertainties are treated separately, for display in
Table~\ref{tab:sys} and in the figures the total experimental systematic uncertainty
in the $\chi_\text{dijet}$ distributions is calculated as the quadratic sum of the
contributions due to the uncertainties in the jet energy
calibration, jet \PT resolution, unfolding correction, and pileup.

\begin{table}
\begin{center}
\topcaption{\label{tab:sys} Summary of main experimental and theoretical uncertainties in the normalized \CHI distributions.
Although the change in the \CHI distribution from each uncertainty is taken into account in the statistical analysis,
this table summarizes the uncertainty in just the smallest \CHI bin, for the smallest and largest bins in dijet mass.
The uncertainty in the dijet bin with largest mass is dominated by the statistical experimental contribution, while the
theoretical contribution is dominated by the uncertainty in the NLO QCD scale.}
\begin{tabular}{l|c|c}
Uncertainty & $1.9<\mjj<2.4$\TeV & $\mjj>4.8$\TeV \\
\hline
Statistical & 1.1\% & 26\% \\
Jet energy scale  & 2.2\% & 3.6\% \\
Jet energy resolution (core) & 0.4\% & 1.1\% \\
Jet energy resolution (tails) & 0.6\% & 0.5\% \\
Unfolding, modelling & 0.1\% & 0.8\% \\
Unfolding, detector simulation & 0.3\% & 0.8\% \\
Pileup & 0.2\% & 0.2\% \\
\hline
Total experimental & 2.6\% & 26\% \\
\hline
\\[-12pt]
\strut
NLO QCD scale & { $_{-2.8\%}^{+7.9\%}$} & { $_{-4.9\%}^{+13\%}$} \\[2pt]
PDF (CT14 eigenvectors) & 0.15\% & 0.4\% \\
Nonperturbative effects & $<$1\% & $<$1\% \\
\hline
\\[-12pt]
\strut
Total theoretical & { $_{-2.8\%}^{+7.9\%}$} & { $_{-4.9\%}^{+13\%}$} \\
\end{tabular}
\end{center}
\end{table}

\section{Theoretical prediction and uncertainties}

We compare the measured normalized dijet angular distributions with the
predictions of perturbative QCD at NLO, which are made
with \textsc{nlojet++}~4.1.3~\cite{nlojet} in the \textsc{fastnlo}~2.1
framework~\cite{fastnlo}.  With the inclusion of
the electroweak (EW) corrections for dijet production~\cite{Dittmaier:2012kx},
the predictions of the normalized $\chi_\text{dijet}$
distributions are corrected up to 1\% and up to 5\% at small and
large \mjj, respectively.  The factorization ($\mu_{\mathrm{f}}$) and renormalization
($\mu_{\mathrm{r}}$) scales are set to the average \pt of the two jets, \ptave,
and the PDFs are taken from the CT14 set~\cite{Dulat:2015mca}.
The use of a more flexible statistical combination of multiple PDF
sets as in PDF4LHC15\_100~\cite{Butterworth:2015oua, Dulat:2015mca,
  Harland-Lang:2014zoa, Ball:2014uwa, Gao:2013bia, Carrazza:2015aoa}
exhibited only small differences as compared to the use of the CT14 PDF set alone,
and had negligible impact on the CI limits described in the next section.

We evaluated the impact on the QCD predictions of nonperturbative effects related to
hadronization and multiple parton interactions
using {\PYTHIA{}~8} with the CUETP8M1
tune~\cite{Skands:2014pea,Khachatryan:2015pea} and {\HERWIGpp}
2.7.1~\cite{herwig:2008} with tune EE5C. The effects are found to be
negligible in both MC event generators.  We can therefore compare the
data corrected to particle-level with the parton-level theory
predictions.

The choices of the $\mu_{\mathrm{f}}$ and $\mu_{\mathrm{r}}$ scales dominate the
uncertainties in the QCD prediction.  These uncertainties are evaluated
following the proposal in Refs.~\cite{Cacciari:2003fi,Banfi:2010xy} by changing the
default choice of scales in the following 6 combinations:
$(\mu_{\mathrm{f}}/\ptave$, $\mu_{\mathrm{r}}/\ptave)$ = $(1/2, 1/2)$, $(1/2, 1)$,
$(1,1/2)$, $(2, 2)$, $(2, 1)$, and $(1, 2)$.  These changes
modify the predictions of the normalized $\chi_\text{dijet}$
distributions by up to 8\% and up to 13\% at small and large values of \mjj, respectively.
The uncertainty due to the choice of PDFs is determined from the 28
eigenvectors of CT14 using the procedure described in
Ref.~\cite{cteq}, and is found to be less than 0.15\% at low \mjj
and less than 0.4\% at high \mjj. The uncertainty of the strong
coupling constant has a negligible impact on the normalised
\CHI distribution. A summary of the leading
systematic uncertainties in the theoretical predictions is also given in
Table~\ref{tab:sys}.

New physics signatures from CIs with flavor-diagonal color-singlet couplings between quarks are
studied. These are described by the effective
Lagrangian~\cite{CI-1,CI-2}:
\begin{equation*}
  \mathcal{L}_{\cPq\cPq}=\frac{2\pi}{\Lambda^2} \left[
    \eta_{\mathrm{LL}}({\overline{\cPq}}_{\mathrm{L}}\gamma^{\mu}\cPq_{\mathrm{L}})(\overline{\cPq}_{\mathrm{L}}\gamma_{\mu}\cPq_{\mathrm{L}})
    +\eta_{\mathrm{RR}}(\overline{\cPq}_{\mathrm{R}}\gamma^{\mu}\cPq_{\mathrm{R}})(\overline{\cPq}_{\mathrm{R}}\gamma_{\mu}\cPq_{\mathrm{R}})
    +2\eta_{\mathrm{RL}}(\overline{\cPq}_{\mathrm{R}}\gamma^{\mu}\cPq_{\mathrm{R}})(\overline{\cPq}_{\mathrm{L}}\gamma_{\mu}\cPq_{\mathrm{L}})
  \right],
\end{equation*}
where the subscripts $\mathrm{L}$ and $\mathrm{R}$ refer to the left and right chiral projections of
the quark fields, respectively, and $\eta_{\mathrm{LL}}$, $\eta_{\mathrm{RR}}$, and $\eta_{\mathrm{RL}}$ are
given the values of 0, ${+}1$, or ${-}1$. The various combinations of
$\left(\eta_{\mathrm{LL}},\,\eta_{\mathrm{RR}},\,\eta_{\mathrm{RL}}\right)$ correspond to
different CI models. The following CI possibilities with color-singlet
couplings among quarks are investigated:
\begin{center}
\renewcommand{\arraystretch}{1.14}
  \begin{tabular}{c|c}
    $\Lambda$ & $\left(\eta_{\mathrm{LL}},\,\eta_{\mathrm{RR}},\,\eta_{\mathrm{RL}}\right)$\\\hline
    $\Lambda^{\pm}_{\mathrm{LL}}$    & $(\pm 1,\hphantom{\pm}0,\hphantom{\pm}0)\,$\\
    $\Lambda^{\pm}_{\mathrm{RR}}$    & $(\hphantom{\pm}0,\pm 1,\hphantom{\pm}0)\,$\\
    $\Lambda^{\pm}_{\mathrm{VV}}$    & $(\pm 1,\pm 1,\pm 1)\,$\\
    $\Lambda^{\pm}_{\mathrm{AA}}$    & $(\pm 1,\pm 1,\mp 1)\,$\\
    $\Lambda^{\pm}_{(\mathrm{V-A})}$ & $(\hphantom{\pm}0,\hphantom{\pm}0,\pm 1)\,$\\
  \end{tabular}
  \renewcommand{\arraystretch}{1}
\end{center}
The models with positive (negative) $\eta_{\mathrm{LL}}$ or $\eta_{\mathrm{RR}}$ lead to
destructive (constructive) interference with the QCD terms, and a lower (higher) cross section, respectively.
In all CI models discussed in this paper, NLO QCD corrections are
employed to calculate the cross sections. In proton-proton collisions the $\Lambda^{\pm}_{\mathrm{LL}}$ and
$\Lambda^{\pm}_{\mathrm{RR}}$ models result in identical tree level cross sections and NLO corrections,
and consequently lead to the same sensitivity. For
$\Lambda^{\pm}_{\mathrm{VV}}$ and $\Lambda^{\pm}_{\mathrm{AA}}$, as well as for
$\Lambda^{\pm}_{(\mathrm{V-A})}$, the CI predictions are
identical at tree level, but exhibit different NLO corrections and yield different sensitivity.
For calculating the CI terms, as well as the interference between the
CI terms and QCD terms at leading order (LO) and NLO in QCD, the \textsc{cijet}~1.0 program~\cite{contactnlo} is employed.

For the ADD model, two parameterizations for virtual graviton exchange are considered,
Giudice--Rattazzi--Wells (GRW)~\cite{GRW} and Han--Lykken--Zhang
(HLZ)~\cite{HLZ}.
In the GRW convention, the sum over the Kaluza--Klein
graviton excitations in the effective field theory is regulated by a
single cutoff parameter $\Lambda_{\mathrm{T}}$.
In the HLZ convention, the effective
theory is described in terms of two parameters, the cutoff scale $M_{\mathrm{S}}$ and the number
of extra spatial dimensions $n_{\mathrm{ED}}$.
The parameters $M_{\mathrm{S}}$ and $n_{\mathrm{ED}}$ are directly related to
$\Lambda_{\mathrm{T}}$~\cite{landsberg}.  We consider models with 2--6 EDs.
The case of $n_{\mathrm{ED}}=1$ is not considered since it would require an
ED of the size of the order of the solar system; the gravitational potential
at these distances would be noticeably modified and this case is therefore excluded by observation.
The case of $n_{\mathrm{ED}}=2$ is special in
the sense that the relation between $M_{\mathrm{S}}$ and $\Lambda_{\mathrm{T}}$ also
depends on the parton-parton center-of-mass energy $\sqrt{\hat{s}}$.
The ADD predictions are calculated with {\PYTHIA{}8}.

Quantum black hole production is studied within the framework of the ADD model with $n_{\mathrm{ED}}=6$,
and the Randall--Sundrum model with $n_{\mathrm{ED}}=1$ (RS1)~\cite{RS1,RS2}.
In these models, the QBH production cross section
is typically described by the classical geometrical cross section $\sigma_\mathrm{QBH} \approx \pi r_{\mathrm{s}}^{2}$,
where $r_{\mathrm{s}}$ is the Schwarzschild radius of the black hole. The Schwarzschild radius depends on
the mass of the QBH, the Planck scale ($M_{\mathrm{P}}$), and the number of spatial dimensions.
Since QBHs are produced with mass threshold close to the Planck scale, we set the minimum quantum black hole mass
$M_{\mathrm{QBH}}$ equal to $M_{\mathrm{P}}$ for simplicity.
The \QBHMC 3.0 generator~\cite{QBHGenRef} is used for the predictions.

To take into account the NLO QCD and EW corrections to SM dijet production when probing the ADD and QBH
models, the cross section difference
$\sigma^{\mathrm{QCD}}_{\text{NLO+EW corr}} - \sigma^{\mathrm{QCD}}_{\text{LO}}$
is evaluated for each \mjj and \CHI bin and added to the ADD and QBH
predictions. This procedure provides an SM+ADD or SM+QBH prediction wherein
the QCD terms are corrected to NLO with EW corrections while the ADD or QBH
terms are calculated at LO.
In all the predictions, changes from theoretical uncertainties associated with scales
and PDFs are applied only to the QCD prediction, thereby treating the effective NP terms as fixed benchmark terms.

\section{Results}

\begin{figure}
\centering
\includegraphics[width=0.7\textwidth]{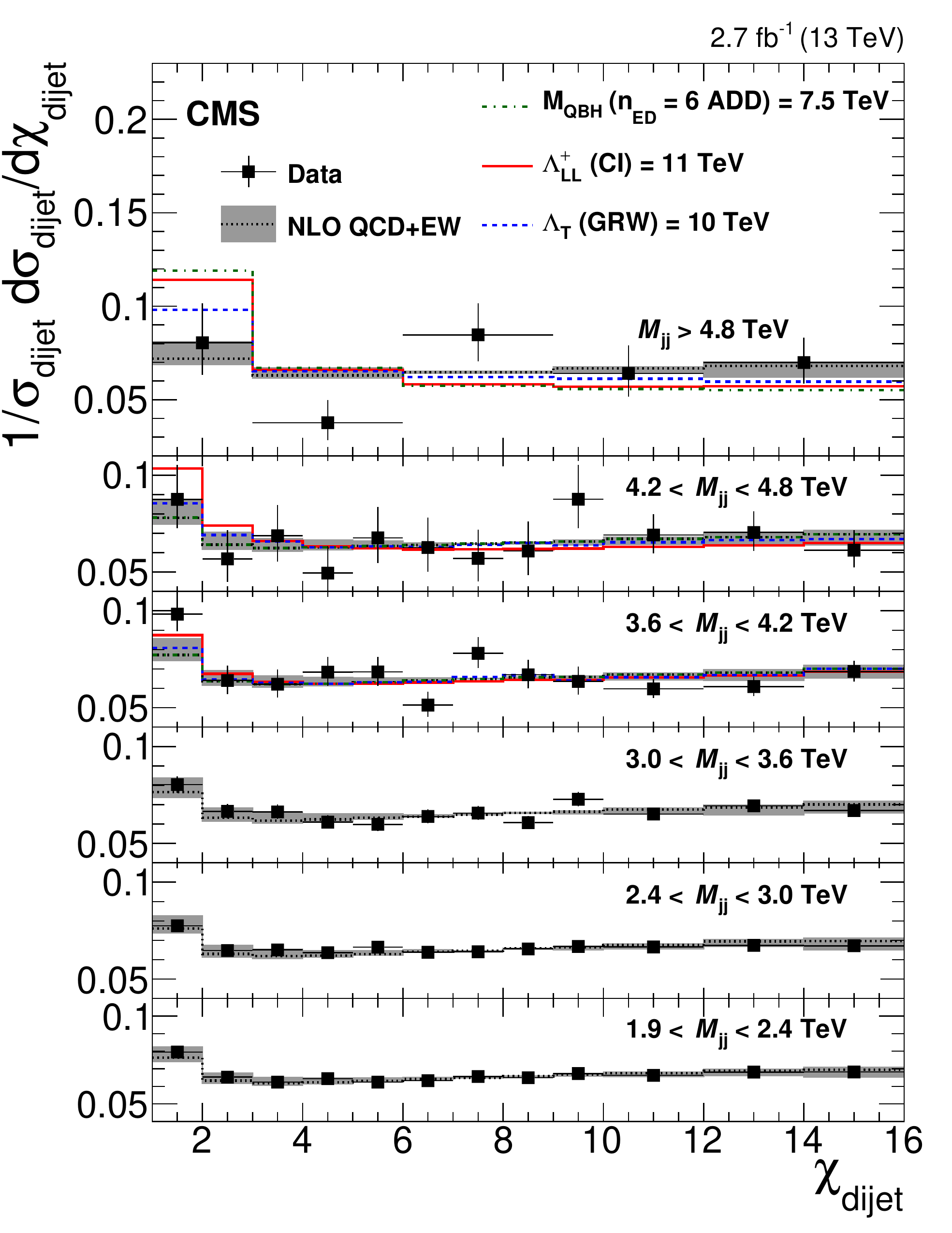}
\caption{Normalized \CHI distributions for \intlumi of integrated luminosity at $\sqrt{s} = 13$\TeV. The corrected
 distributions in data are compared to NLO predictions (black dotted line).
 The vertical bar on each data point represents statistical and systematic experimental uncertainties combined in quadrature.
 The horizontal bar indicates the bin width.
Theoretical uncertainties are indicated by the gray bands. Also shown are the predictions for
QCD+QBH with $n_{\mathrm{ED}}=6$ and $M_{\mathrm{QBH}}=7.5$\TeV (green dashed-dotted line),
QCD+CI with $\Lambda_{\mathrm{LL}}^{+}=11$\TeV (red solid line),
and QCD+ADD with $\Lambda_{\mathrm{T}}\,(\mathrm{GRW})=10$\TeV (blue dashed line).
 }
\label{fig:data_results}
\end{figure}

\begin{figure}
\centering
\includegraphics[width=0.49\textwidth]{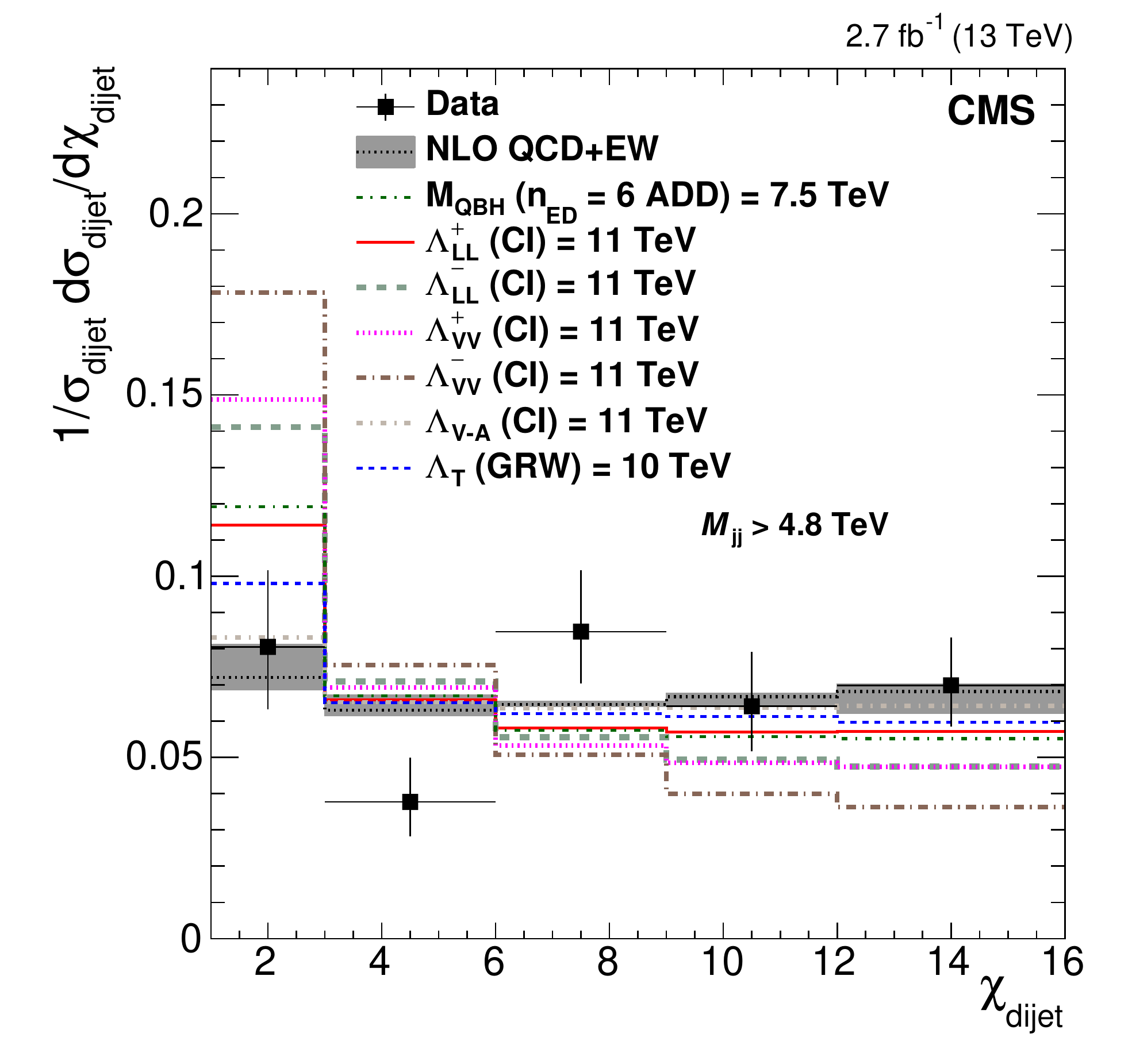}
\includegraphics[width=0.49\textwidth]{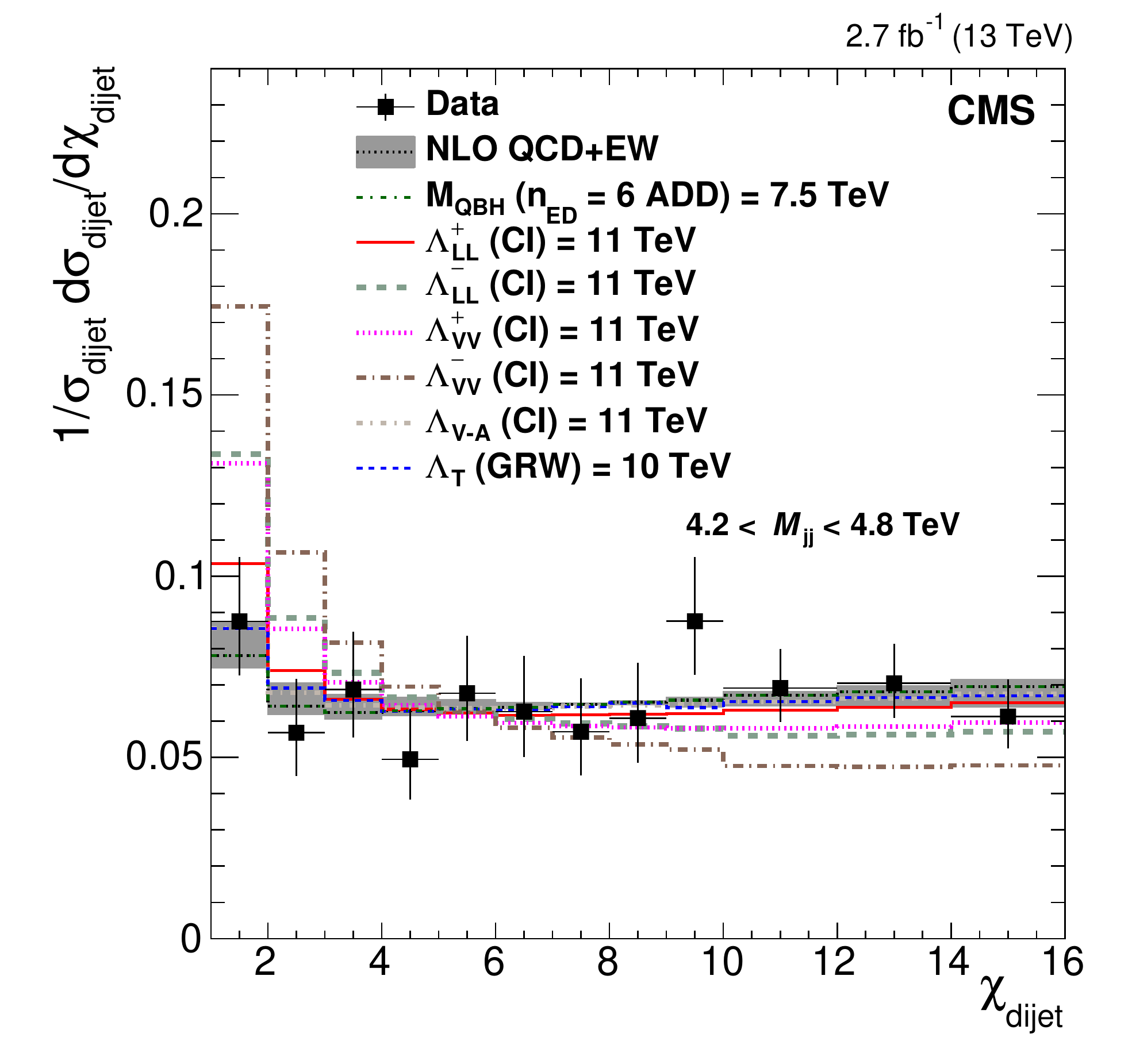}
\includegraphics[width=0.49\textwidth]{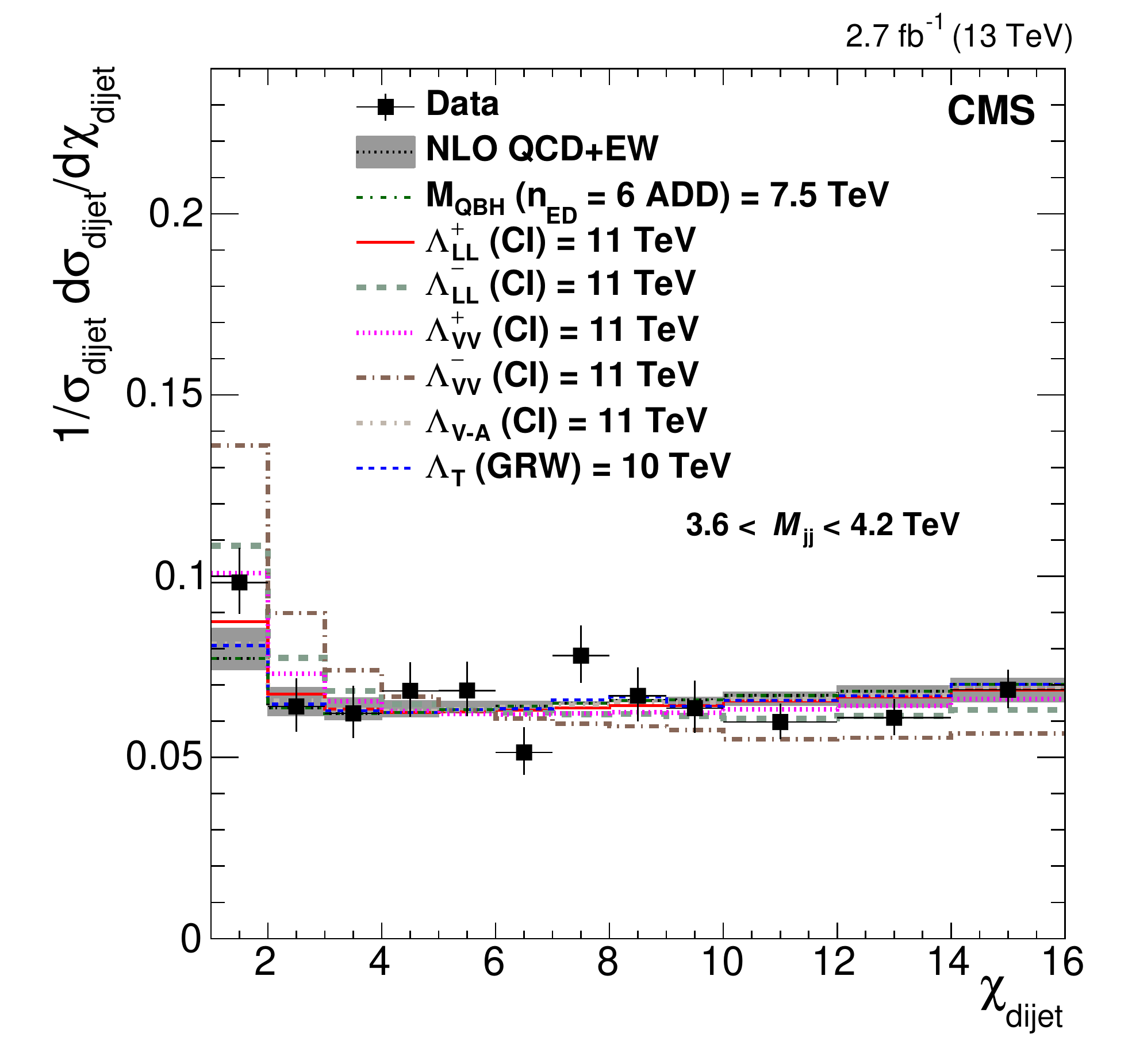}
\caption{Normalized \CHI distributions for \intlumi of integrated luminosity in the highest three mass bins. The corrected
 distributions in data are compared to NLO predictions with non-perturbative corrections (black dotted line).
The vertical bar on each data point represents statistical and systematic experimental uncertainties combined in quadrature.
 The horizontal bar indicates the bin width. Theoretical uncertainties are indicated by the gray band. Also shown are the predictions for
 various QBH, CI, and ADD models.
 }
\label{fig:data_results2}
\end{figure}

The normalized \CHI distributions for all mass bins are compared to NLO predictions with
EW corrections in Fig.~\ref{fig:data_results}. No significant deviation from the theory
is observed. The distributions are also compared to predictions for QCD+CI with $\Lambda_{\text{LL}}^{+}=11$\TeV,
QCD+ADD with $\Lambda_{\mathrm{T}}\ (\mathrm{GRW})\ =10\TeV$, and QCD+QBH with $M_{\mathrm{QBH}}\,(n_{\mathrm{ED}}=6\,\mathrm{ADD})=7.5\TeV$.
The QCD+ADD $\Lambda_{\mathrm{T}}\,(\mathrm{GRW}) =10\TeV$ prediction corresponds to QCD+ADD $M_{\mathrm{S}}\,(\mathrm{HLZ})=10.1$, 11.9, 10.0, 9.9 and 8.4\TeV for $n_{\mathrm{ED}}=2$, 3, 4, 5 and 6, respectively.
The signal distributions are shown only for the highest three ranges of \mjj, since those bins dominate the sensitivity to the NP signals considered.
An expanded version of the normalized \CHI distributions in the highest three ranges of \mjj is shown in Fig.~\ref{fig:data_results2}.
The measured \CHI distributions are used to determine exclusion limits on the NP models.

A modified frequentist approach~\cite{junk:1999,read:2002}
is used to set exclusion limits on the scale $\Lambda$.
The log-likelihoods $L_\mathrm{QCD}$ and $L_\mathrm{QCD+NP}$ are defined
for the respective QCD-only and QCD+NP hypotheses
as a product of Poissonian likelihood functions for each bin in
$\chi_\text{dijet}$ for the highest three ranges of \mjj.
The predictions for each \mjj range are normalized to the number
of observed events in that range.
The $p$-values for the two hypotheses,
$P_\mathrm{QCD+NP}(q \geq q_\text{obs})$ and $P_\mathrm{QCD}(q \leq q_\text{obs})$,
are based on the log-likelihood ratio $q = -2 \ln ({L_\text{QCD+NP}}/{L_\mathrm{QCD}})$.
They are evaluated by generating distributions of $q$ using ensembles of pseudo-experiments,
where systematic uncertainties are represented as Gaussian-constraint
nuisance parameters and are treated according to the frequentist paradigm~\cite{cousinshighland}.
Limits on the QCD+NP models are set based on the quantity
$\mathrm{CL}_\mathrm{s} = P_\mathrm{QCD+NP}(q\geq q_\text{obs}) /
(1-P_\mathrm{QCD}(q \leq q_\text{obs}))$, which is required to be
less than 0.05 for an exclusion at 95\% confidence level (CL).
The observed and
expected exclusion limits on different CI, ADD, and QBH models obtained
in this analysis at 95\% CL are listed in Table~\ref{tab:limits}.
The observed limits are smaller than the expected limits owing to a
slight excess of events in the lowest \CHI bin in the 3.6--4.2\TeV mass bin.
The limits on $M_{\mathrm{S}}$ for the different
numbers of extra dimensions, $n_{\mathrm{ED}}$, directly follow from the limit
for $\Lambda_{\mathrm{T}}$. The limits for the CI scale
$\Lambda_{\mathrm{LL/RR}}^{+}$ are also determined for the case in which the
data are not corrected for detector effects, and are found to agree
with the quoted ones within 3\%.

\begin{table}
\centering
\caption{Observed and expected exclusion limits at 95\% CL for various CI, ADD, and QBH models.}
\label{tab:limits}
\begin{tabular}{l | d{-2} | D{,}{\pm}{-1} }
Model & \multicolumn{1}{c|}{Observed lower limit (TeV)} & \multicolumn{1}{c}{Expected lower limit (TeV)} \\
\cline{1-3}
$\Lambda_{\mathrm{LL/RR}}^{+}$ (NLO) & 11.5 & 12.1 , 1.2\\
$\Lambda_{\mathrm{LL/RR}}^{-}$ (NLO) & 14.7 & 17.3 , 3.4\\
$\Lambda_{\mathrm{VV}}^{+}$ (NLO) & 13.3 & 13.9 , 1.2\\
$\Lambda_{\mathrm{VV}}^{-}$ (NLO) & 18.6 & 22.2 , 5.4\\
$\Lambda_{\mathrm{AA}}^{+}$ (NLO) & 13.3 & 13.9 , 1.2\\
$\Lambda_{\mathrm{AA}}^{-}$ (NLO) & 18.6 & 22.1 , 5.1\\
$\Lambda_{(\mathrm{V-A})}^{+}$ (NLO) & 8.4 & 9.5 , 1.6\\
$\Lambda_{(\mathrm{V-A})}^{-}$ (NLO) & 8.4 & 9.5 , 1.7\\
\cline{1-3}
ADD $\Lambda_{\mathrm{T}}$ (GRW) & 9.4 & 9.8 , 1.2 \\
ADD $M_{\mathrm{S}}$ (HLZ) $n_{\mathrm{ED}}=2$ & 10.1 & 10.6 , 1.3 \\
ADD $M_{\mathrm{S}}$ (HLZ) $n_{\mathrm{ED}}=3$ & 11.2 & 11.7 , 1.4 \\
ADD $M_{\mathrm{S}}$ (HLZ) $n_{\mathrm{ED}}=4$ & 9.4 & 9.8 , 1.2 \\
ADD $M_{\mathrm{S}}$ (HLZ) $n_{\mathrm{ED}}=5$ & 8.5 & 8.9 , 1.1 \\
ADD $M_{\mathrm{S}}$ (HLZ) $n_{\mathrm{ED}}=6$ & 7.9 & 8.2 , 1.0 \\
\hline
$n_{\mathrm{ED}}=6$ ADD QBH $M_{\mathrm{QBH}}$ & 7.8 & 7.7 , 0.3\\
$n_{\mathrm{ED}}=1$ RS QBH $M_{\mathrm{QBH}}$  & 5.3 & 5.3 , 0.4\\
\end{tabular}
\end{table}

The agreement of the data with QCD predictions is quantified by
calculating $P_\mathrm{QCD}(q \leq q_\text{obs})$ as described above.
The largest excess is found in the 3.6--4.2\TeV mass bin with a
significance of 1.8 standard deviations.

\section{Summary}

Normalized dijet angular distributions have been measured at $\sqrt{s} = 13$\TeV
with the CMS detector over a wide range of dijet invariant masses. The
distributions are found to be in agreement with predictions
of perturbative QCD and are used to set lower limits on the contact-interaction scale for a
variety of quark-compositeness models that include next-to-leading order QCD corrections,
models with large extra dimensions, and models of quantum black-hole production. The 95\% confidence level lower limits for the
contact interaction scale $\Lambda$ are in the range 8.4--18.6\TeV.
Also excluded are quantum black holes with masses up to 7.8 \TeV
in the ADD model for $n_{\mathrm{ED}} = 6$, and up to 5.3\TeV in the
Randall--Sundrum model for $n_{\mathrm{ED}}=1$.
The lower limits for the scales of ADD models, $\Lambda_{\mathrm{T}}$\,(GRW) and $M_{\mathrm{S}}$\,(HLZ),
are in the range 7.9--11.2\TeV, and are the most stringent set of limits available.

\begin{acknowledgments}
We congratulate our colleagues in the CERN accelerator departments for the excellent performance of the LHC and thank the technical and administrative staffs at CERN and at other CMS institutes for their contributions to the success of the CMS effort. In addition, we gratefully acknowledge the computing centers and personnel of the Worldwide LHC Computing Grid for delivering so effectively the computing infrastructure essential to our analyses. Finally, we acknowledge the enduring support for the construction and operation of the LHC and the CMS detector provided by the following funding agencies: BMWFW and FWF (Austria); FNRS and FWO (Belgium); CNPq, CAPES, FAPERJ, and FAPESP (Brazil); MES (Bulgaria); CERN; CAS, MoST, and NSFC (China); COLCIENCIAS (Colombia); MSES and CSF (Croatia); RPF (Cyprus); SENESCYT (Ecuador); MoER, ERC IUT, and ERDF (Estonia); Academy of Finland, MEC, and HIP (Finland); CEA and CNRS/IN2P3 (France); BMBF, DFG, and HGF (Germany); GSRT (Greece); OTKA and NIH (Hungary); DAE and DST (India); IPM (Iran); SFI (Ireland); INFN (Italy); MSIP and NRF (Republic of Korea); LAS (Lithuania); MOE and UM (Malaysia); BUAP, CINVESTAV, CONACYT, LNS, SEP, and UASLP-FAI (Mexico); MBIE (New Zealand); PAEC (Pakistan); MSHE and NSC (Poland); FCT (Portugal); JINR (Dubna); MON, RosAtom, RAS, RFBR and RAEP (Russia); MESTD (Serbia); SEIDI, CPAN, PCTI and FEDER (Spain); Swiss Funding Agencies (Switzerland); MST (Taipei); ThEPCenter, IPST, STAR, and NSTDA (Thailand); TUBITAK and TAEK (Turkey); NASU and SFFR (Ukraine); STFC (United Kingdom); DOE and NSF (USA).

\hyphenation{Rachada-pisek} Individuals have received support from the Marie-Curie program and the European Research Council and EPLANET (European Union); the Leventis Foundation; the A. P. Sloan Foundation; the Alexander von Humboldt Foundation; the Belgian Federal Science Policy Office; the Fonds pour la Formation \`a la Recherche dans l'Industrie et dans l'Agriculture (FRIA-Belgium); the Agentschap voor Innovatie door Wetenschap en Technologie (IWT-Belgium); the Ministry of Education, Youth and Sports (MEYS) of the Czech Republic; the Council of Science and Industrial Research, India; the HOMING PLUS program of the Foundation for Polish Science, cofinanced from European Union, Regional Development Fund, the Mobility Plus program of the Ministry of Science and Higher Education, the National Science Center (Poland), contracts Harmonia 2014/14/M/ST2/00428, Opus 2014/13/B/ST2/02543, 2014/15/B/ST2/03998, and 2015/19/B/ST2/02861, Sonata-bis 2012/07/E/ST2/01406; the National Priorities Research Program by Qatar National Research Fund; the Programa Clar\'in-COFUND del Principado de Asturias; the Thalis and Aristeia programs cofinanced by EU-ESF and the Greek NSRF; the Rachadapisek Sompot Fund for Postdoctoral Fellowship, Chulalongkorn University and the Chulalongkorn Academic into Its 2nd Century Project Advancement Project (Thailand); and the Welch Foundation, contract C-1845. \end{acknowledgments}

\bibliography{auto_generated}
\cleardoublepage \appendix\section{The CMS Collaboration \label{app:collab}}\begin{sloppypar}\hyphenpenalty=5000\widowpenalty=500\clubpenalty=5000\input{EXO-15-009-authorlist.tex}\end{sloppypar}
\end{document}

%% file: EXO-15-009-authorlist.tex
\textbf{Yerevan Physics Institute,  Yerevan,  Armenia}\\*[0pt]
A.M.~Sirunyan, A.~Tumasyan
\vskip\cmsinstskip
\textbf{Institut f\"{u}r Hochenergiephysik,  Wien,  Austria}\\*[0pt]
W.~Adam, E.~Asilar, T.~Bergauer, J.~Brandstetter, E.~Brondolin, M.~Dragicevic, J.~Er\"{o}, M.~Flechl, M.~Friedl, R.~Fr\"{u}hwirth\cmsAuthorMark{1}, V.M.~Ghete, C.~Hartl, N.~H\"{o}rmann, J.~Hrubec, M.~Jeitler\cmsAuthorMark{1}, A.~K\"{o}nig, I.~Kr\"{a}tschmer, D.~Liko, T.~Matsushita, I.~Mikulec, D.~Rabady, N.~Rad, B.~Rahbaran, H.~Rohringer, J.~Schieck\cmsAuthorMark{1}, J.~Strauss, W.~Waltenberger, C.-E.~Wulz\cmsAuthorMark{1}
% ...
\vskip\cmsinstskip
% ...
1:~~Also at Vienna University of Technology, Vienna, Austria\\
% ...
